\documentclass[twocolumn]{webofc}

\usepackage[varg]{txfonts}   
\usepackage{multirow}
\begin{document}
\title{Coexistence and evolution of shapes: mean-field-based interacting boson model}
%
%

\author{\firstname{Kosuke} \lastname{Nomura}\inst{1}\fnsep\thanks{\email{knomura@phy.hr}} }

\institute{Department of Physics, Faculty of Science, University of Zagreb, 10000 Zagreb, Croatia}

\abstract{%
A method of deriving the Hamiltonian of the interacting boson model, that is based on the microscopic framework of the nuclear energy density functional, is presented. The constrained self-consistent mean-field calculation with a given energy density functional provides potential energy surface within the relevant collective coordinates, which is subsequently mapped onto the expectation value of the interacting-boson Hamiltonian in the boson condensate state. This procedure completely determines the strength parameters of the IBM, and the diagonalization of the mapped Hamiltonian yields excitation spectra and transition rates for a given nucleus. Two recent applications of the method are discussed, that is, the descriptions of  the intruder states in Cadmium isotopes and the octupole correlations in neutron-rich odd-mass Barium isotopes. 
}
\maketitle
\section{Introduction}
\label{intro}
The nuclear shapes and collective excitations have been one of the most prominent and studied themes of nuclear structure physics. Experiments using radioactive-ion beams allow to study exotic nuclei that exhibit variety of shapes and related excitations, e.g., evolution and coexistence of quadrupole deformed shapes, pear-shaped (or octupole) deformation, etc., and necessitate timely systematic, as well as reliable, theoretical investigations. 

The interacting boson model (IBM) \cite{IBM}, a model where correlated $J=0^+$ ($S$) and $2^+$ ($D$) nucleon pairs are approximated to ($s$ and $d$) bosons, has been remarkably successful in phenomenological description of low-lying states in nuclei. On the other hand, microscopic foundation of the IBM, i.e., an attempt to derive the bosonic Hamiltonian from nucleonic degrees of freedom, has been extensively pursued. Conventionally, this has been done  within the seniority scheme of the nuclear shell model \cite{OAI}, but this approach has been somewhat limited to nearly spherical nuclei, due to the fact that in deformed systems the seniority classification breaks down in the presence of non-degenerate shells. 

In this context, a completely different method of deriving the Hamiltonian of the IBM was developed \cite{nomura2008}: the results of the self-consistent mean-field (SCMF) calculation within the framework of the nuclear density functional theory (DFT) are used as a microscopic input to determine the Hamiltonian of the IBM. 
Since the DFT framework allows for a global mean-field description of many nuclear properties over the entire region of the nuclear chart, it has become possible to derive the IBM Hamiltonian for any arbitrary nuclei with a variety of shapes: strongly axially-deformed  \cite{nomura2011rot}, $\gamma$-soft \cite{nomura2012tri}, and octupole \cite{nomura2014} shapes, and shape coexistence \cite{nomura2016sc}. This facilitates a detailed, computationally feasible, and systematic prediction of spectroscopy in nuclei, including those far from the stability.  
%
Here we describe the basic notions of this method and highlight two of its recent applications: the descriptions of intruder states in even-even Cd isotopes \cite{nomura2018cd}, and the onset of octupole deformation in neutron-rich odd-mass Ba isotopes \cite{nomura2018oct}.


\section{Mean-field derivation of the IBM}
\label{DFT-IBM}

\begin{figure}[htb!]
\begin{center}
\includegraphics[width=\columnwidth]{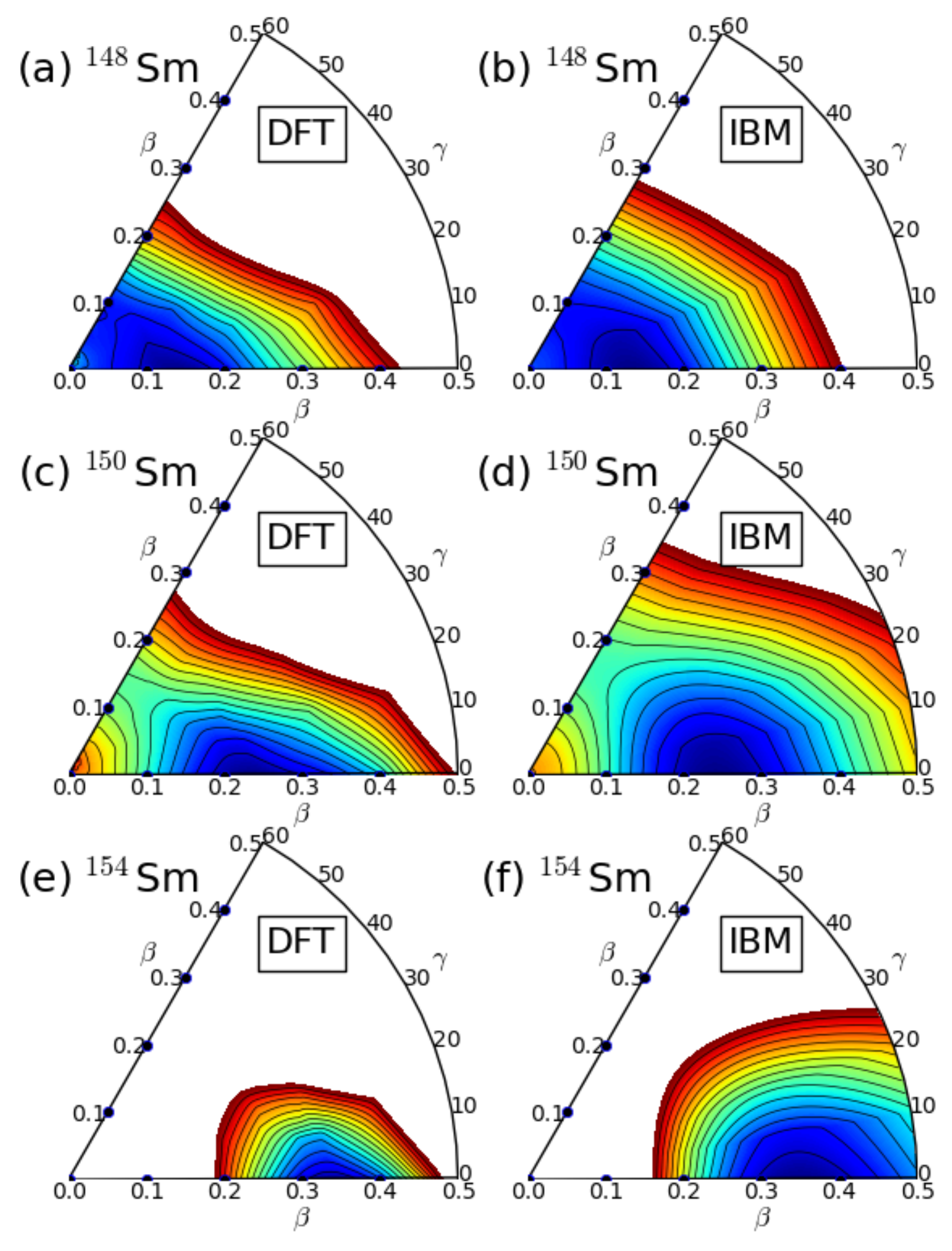}
\caption{(Color online) The SCMF and mapped IBM PESs for the $^{148,152,154}$Sm isotopes. The Skyrme-SkM* functional and the density-dependent zero-range pairing force with strength 1250 MeV fm$^3$ were used. Energy difference between neighbouring contours is 200 keV.} 
\label{fig:sm-pes}
\end{center}
\end{figure}

The first step is a standard constrained SCMF calculation based on a given relativistic \cite{vretenar2005} or non-relativistic \cite{bender2003} EDF. 
In this case, the constraints are on mass
quadrupole moments related to the axial deformation $\beta$ and
non-axial deformation $\gamma$ in the geometrical collective model \cite{BM}. 
The SCMF calculations are performed to obtain a potential energy surface (PES) for a given set of collective coordinates $(\beta,\gamma)$. 
Examples in Fig.~\ref{fig:sm-pes} show the SCMF PESs for the axially deformed Sm nuclei, which are obtained from the constrained Hartree-Fock+BCS method. 
The PES for $^{148}$Sm exhibits a nearly spherical minimum, typical of the vibrational nucleus. It becomes soft in $\beta$ and $\gamma$ in $^{150}$Sm, which is then considered a transitional nucleus. Finally, on the PES for $^{154}$Sm one sees well defined prolate minimum, indicating that this nucleus is strongly deformed. 

Having the SCMF PES for a given deformation $(\beta,\gamma)$, denoted by $E_\mathrm{SCMF}(\beta,\gamma)$,  
for each nucleus,
we subsequently map it to the IBM PES, $E_\mathrm{IBM}(\beta,\gamma)$. 
To be more specific, the IBM Hamiltonian is constructed, i.e., the
IBM strengths parameters are determined, so as to reproduce the SCMF surface at each set of ($\beta,\gamma$) as close as possible. 

We consider the version of the IBM, that distinguishes between neutrons and protons (IBM-2) \cite{OAI}. In the IBM-2, number of neutron (proton) bosons $N_\nu$ ($N_\pi$) equals the number of pairs of valence neutrons (protons). 
The IBM PES is obtained as the expectation value of a given IBM Hamiltonian in the boson coherent state \cite{ginocchio1980}. 
The coherent state is written as the product of neutron $\lambda_\nu$ 
and proton $\lambda_\pi$ boson condensates
$(N_{\nu}!N_{\pi}!)^{-1/2}(\lambda_{\nu})^{N_{\nu}}(\lambda_{\pi})^{N_{\pi}}|0\rangle$, 
where $\lambda_\tau^\dagger=s_\tau^\dagger+\beta_\tau\cos{\gamma_\tau}d_{\tau,0}^\dagger+\beta_\tau\sin{\gamma_\tau}(d^\dagger_{\tau,+2}+d^\dagger_{\tau,-2})/\sqrt{2}$ ($\tau=\nu$ or $\pi$) and $|0\rangle$ is the inert core. 
The above expression invokes deformation variables for both neutrons ($\beta_{\nu},\gamma_{\nu}$) and protons ($\beta_{\pi},\gamma_{\pi}$). 
The four deformation parameters 
($\beta_{\nu},\beta_{\pi},\gamma_{\nu},\gamma_{\pi}$) could in principle
vary independently but, in 
realistic cases, introducing all of these variables 
entails too much complexity. 
Therefore, and as is usual in the collective model \cite{BM}, we assume equal deformations of the proton and neutron
systems, i.e.,  
$\beta_{\nu}=\beta_{\pi}\equiv\beta_\mathrm{B}$ and
$\gamma_{\nu}=\gamma_{\pi}\equiv\gamma_\mathrm{B}$. 
In addition, the variable $\beta_\mathrm{B}$ in the boson system is proportional to $\beta$
deformation in the collective model, $\beta_\mathrm{B}=C\beta$ with $C$ being coefficient, while $\gamma_\mathrm{B}$ and $\gamma$ are supposed to be identical $\gamma_\mathrm{B}=\gamma$.

Under these conditions, we equate the SCMF to the IBM PESs: 
\begin{equation}
\label{eq:hmap}
E_\mathrm{SCMF}(\beta,\gamma)
\sim
E_\mathrm{IBM}(\beta,\gamma). 
\end{equation}
Note that Eq.~(\ref{eq:hmap}) represents an approximate equality as it
is fulfilled within a limited range of ($\beta,\gamma$) plane, i.e.,
in the vicinity of the global minimum. 
We restrict ourselves to this range because it is most relevant to the low-energy
collective states. 
On the other hand, one should not try to reproduce the region far from the energy
minimum as the topology of the
microscopic energy surface around the region is determined by single-nucleon 
configurations that are, by construction, outside the model space of the
IBM consisting of only collective pairs of valence nucleons. 
For this reason, the IBM PES is, as shown on the right column of Fig.~\ref{fig:sm-pes}, always flat in the
region far from the global minimum. 
The topology of the SCMF PES around global minimum should reflect essential
features of fermion many-body systems, such as the Pauli principle, the antisymmetrization, and
the underlying nuclear forces. 
Through the mapping procedure, these effects are supposed to be incorporated in the IBM.

The IBM Hamiltonian that embodies the essentials of the underlying nucleonic interactions reads: 
\begin{eqnarray}
\label{eq:ham-sg}
 \hat H_\mathrm{B}=\epsilon\hat n_d+\kappa\hat Q_{\nu}\cdot\hat Q_{\pi}+\kappa^{\prime}\hat L\cdot\hat L, 
\end{eqnarray}
where the definitions of each term are found, e.g., in Refs.~\cite{nomura2008,nomura2011rot}. 
The parameters $\epsilon$, $\kappa$, 
$\chi_{\nu}$, and $\chi_{\pi}$ (which appear in the quadrupole operators $\hat Q_{\nu}$ and $\hat Q_{\pi}$), 
plus the coefficient $C$, are to be fitted to the fermionic PES. 
Only the parameter $\kappa'$ should be determined independently of the other parameters in such a way \cite{nomura2011rot}, that a rotational response of nucleonic system, i.e., cranking moment of inertia that is obtained from the SCMF calculation at the equilibrium minimum, be reproduced in the boson system.

\begin{figure}[htb!]
\begin{center}
\includegraphics[width=\columnwidth]{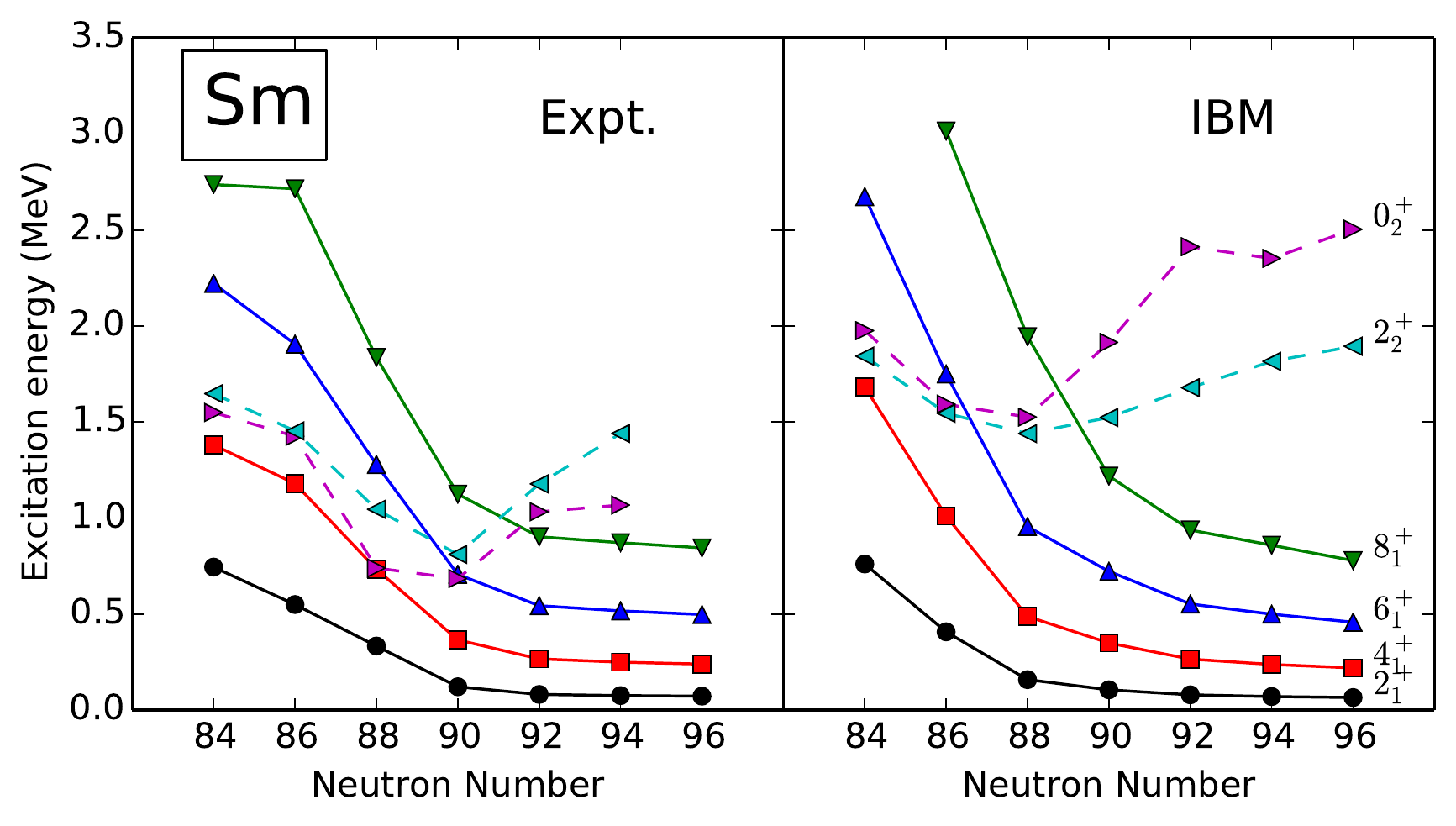}
\caption{(Color online) Experimental and theoretical low-energy levels in Sm isotopes as functions of the neutron number.} 
\label{fig:sm-level}
\end{center}
\end{figure}

Having determined its strength parameters by the above procedure, the resultant IBM Hamiltonian is diagonalized in the laboratory frame, which provides excitation energies and electromagnetic transition rates for a given nucleus. 
The comparison of the calculated low-lying spectra to the experimental ones in Fig.~\ref{fig:sm-level} indicates that the mapped IBM has nicely reproduced the evolution of nuclear structure against the neutron number from nearly spherical vibrational to strongly deformed states.

\section{Structure of even-even Cd isotopes}

The even-even Cd isotopes have been traditionally a well-established example for vibrational nuclei. Experiments carried out more recently, however, identified several additional $0^+$ and $2^+$ at low energy, and the question as to whether these states arise from intruder excitation from other major shell has been a long-standing problem. It is then the purpose of Ref.~\cite{nomura2018cd} that we addressed this question within the SCMF-IBM calculation that includes intruder configuration in the boson system.

\subsection{IBM with configuration mixing}

\begin{figure}[htb!]
\begin{center}
\includegraphics[width=\columnwidth]{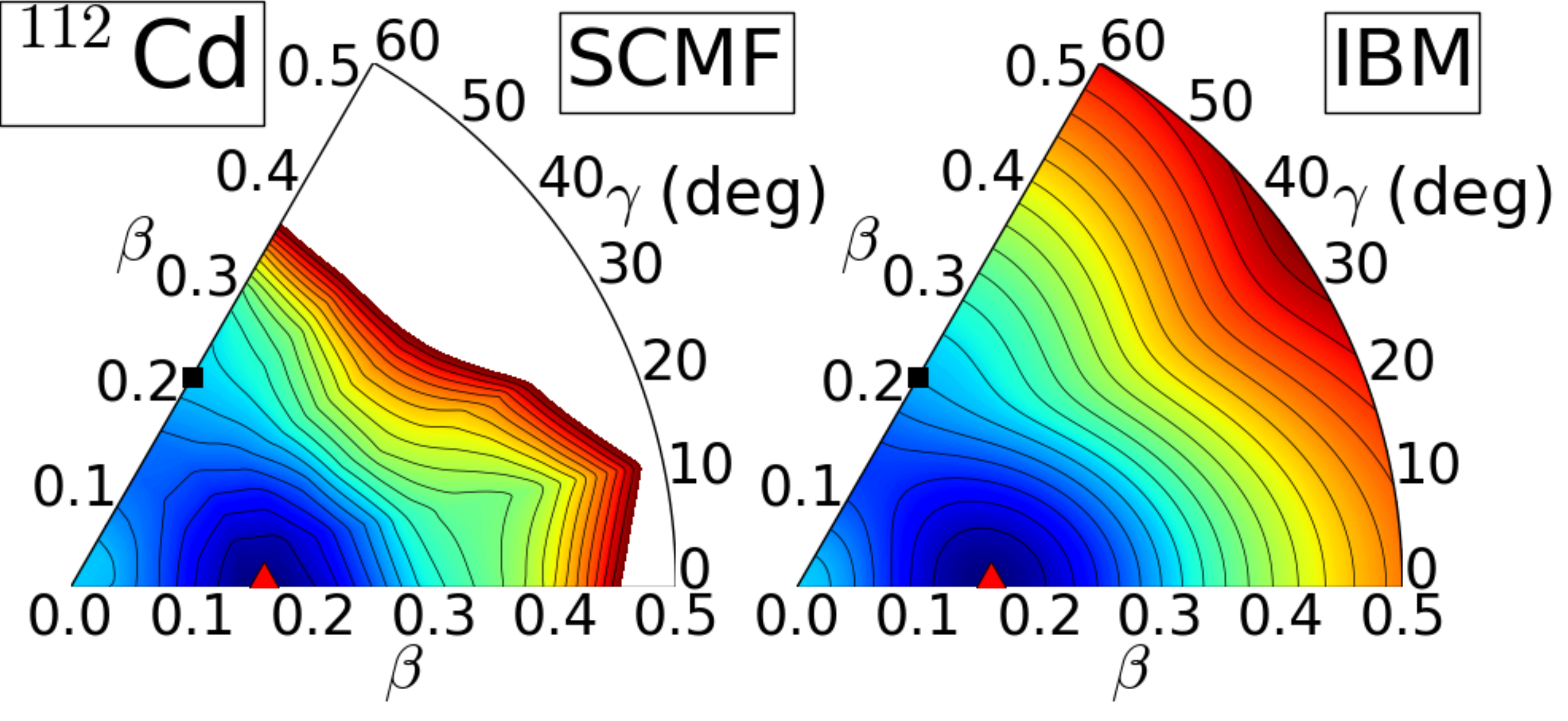}
\caption{(Color online) Left: The $(\beta,\gamma)$ PES for $^{112}$Cd, obtained from the SCMF calculation using
 the Skyrme SLy6 EDF with the density-dependent zero-range pairing
 interaction with the strength 1000 MeV fm$^{3}$. Right: the corresponding IBM-2 PES. Energy difference between neighbouring contours is 250
 keV. The global minimum is indicated by solid triangle, while the local
 minimum is identified by solid squares.} 
\label{fig:pes-cd}
\end{center}
\end{figure}

We draw on the left-hand side in Fig.~\ref{fig:pes-cd} the $(\beta,\gamma)$ SCMF PES for the $^{112}$Cd, based on the Skyrme-SLy6 EDF. 
For all the considered Cd nuclei, a prolate global minimum is found with 
moderate axial deformation $\beta\approx 0.15$. 
We also observe on the oblate side ($\gamma\approx 60^{\circ}$) a much
less pronounced local minimum between $\beta=0.2$ and 0.3.

We then construct from the SCMF result the IBM-2 Hamiltonian that includes configuration mixing. 
A method to incorporate intruder configuration was proposed in Ref.~\cite{duval1981}, that the different shell-model spaces of $0p-0h$, $2p-2h$, $4p-4h$, $\ldots$ excitations be associated with the corresponding boson spaces comprising $N$, $N+2$, $N+4$, $\ldots$ bosons, with $N$ being the total boson number $N=N_{\pi}+N_{\nu}$, and that the different boson subspaces be subsequently mixed. 
In most of the configuration mixing IBM calculations, particle-like and hole-like bosons are not distinguished and, as the excitation of a pair (or boson) increases the boson number by two, the $0p-0h$, $2p-2h$, etc. configurations differ in boson number by two. 
Here we take the doubly-magic nuclei $^{100}$Sn and $^{132}$Sn (for $^{116}$Cd)  as inert cores for the
bosons. Also we take into account up to the $2p-2h$ intruder excitation of protons 
across the $Z=50$ shell closure. Accordingly, $N_{\pi}=1$ and 3 for the normal (i.e., $0p-0h$) and intruder ($2p-2h$) 
configurations, respectively, while $5\leq N_{\nu}\leq 8$. 
The configuration mixing IBM-2 Hamiltonian is then written as: 
\begin{eqnarray}
\label{eq:ham-cm}
 \hat H = {\cal\hat P}_1\hat H_1{\cal\hat P}_1 + {\cal\hat P}_3(\hat H_3 + \Delta){\cal\hat P}_3 + \hat
  H_{\rm mix}, 
\end{eqnarray}
where $\hat H_1$ ($\hat H_3$) and ${\cal\hat P}_1$ (${\cal\hat P}_3$) are the
Hamiltonian of and the projection operator onto the normal and intruder
configuration spaces, respectively. $\Delta$ stands for the energy
needed to promote a proton boson across the shell closure. 
$\hat H_{\rm mix}$ in the above equation is the term that is allowed to
mix the two configurations.

The coherent state for the configuration mixing IBM is defined as the direct sum of the coherent state for each
unperturbed configuration \cite{frank2004}. 
The bosonic PES is obtained as the
lower eigenvalue of the $2\times 2$ coherent-state matrix. 
The parameters for the Hamiltonian for each configuration are determined
by associating the Hamiltonian with each mean-field minimum, i.e., $0p-0h$
Hamiltonian for the prolate minimum, and $2p-2h$ one for the oblate
local minimum, and the energy offset $\Delta$ and the mixing strength in $\hat H_{\rm
mix}$ are determined so that the energy difference between the two
mean-field minima and the barrier height for these minima, respectively,
are reproduced. 
On the right-hand side of Fig.~\ref{fig:pes-cd} we show the mapped IBM-2 PES for $^{112}$Cd. 
One notices that the topology of the corresponding SCMF PES
in the neighborhood of the minimum is well reproduced by the IBM one. 

\subsection{Results for the spectroscopic properties}

\begin{figure*}[htb!]
\begin{center}
\begin{tabular}{ccc}
\includegraphics[width=0.28\linewidth]{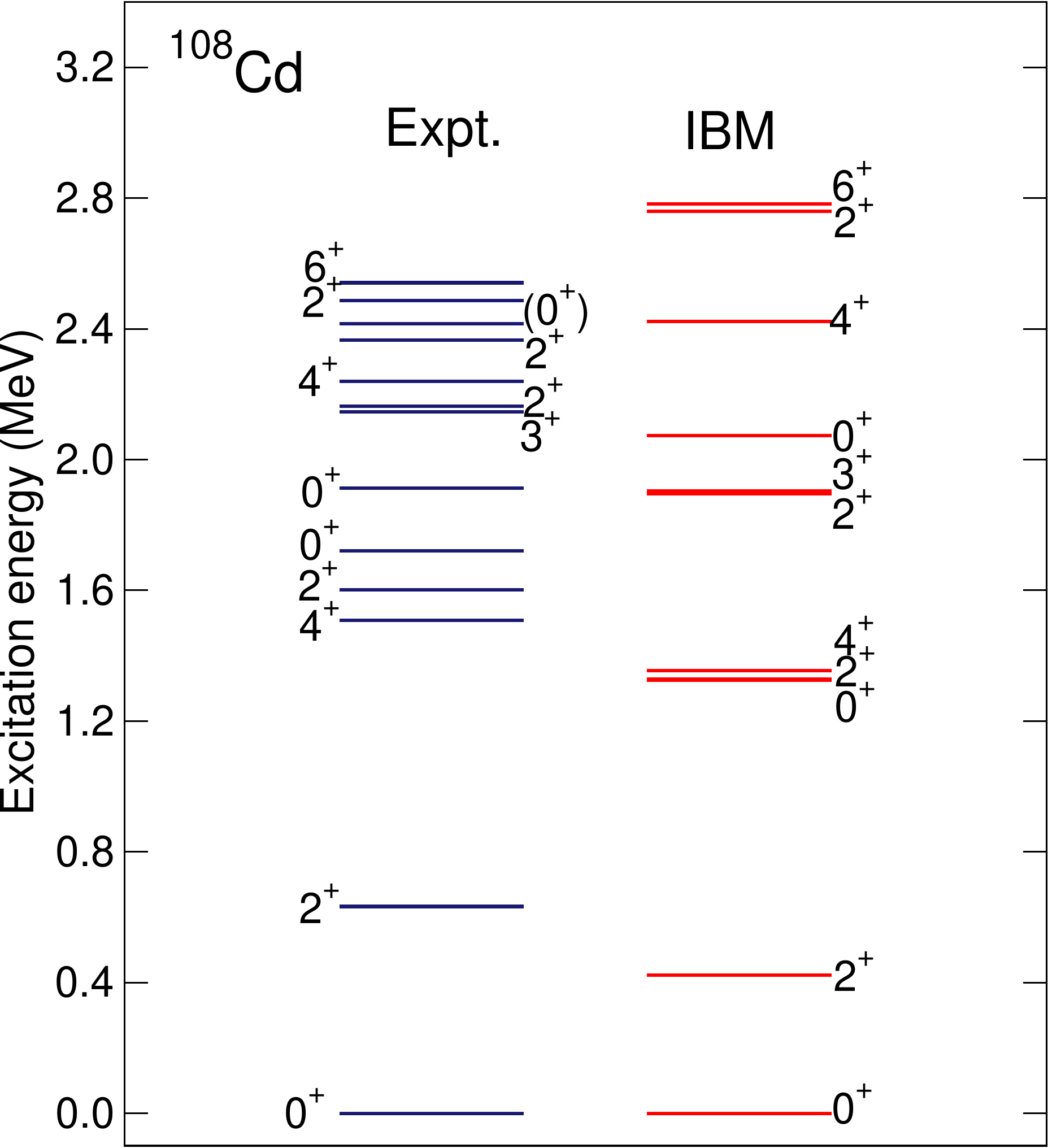} &
\includegraphics[width=0.28\linewidth]{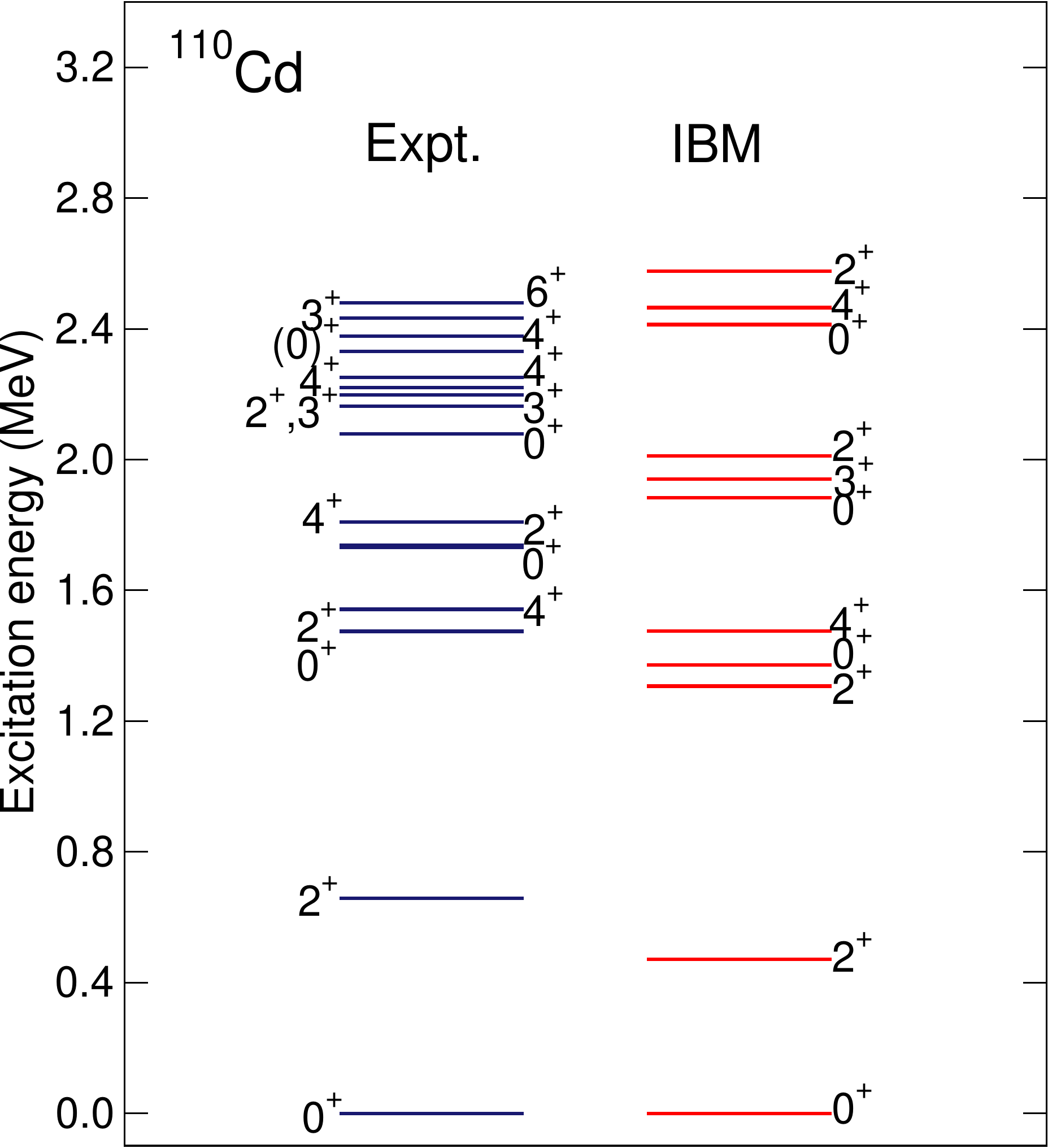} &
\includegraphics[width=0.28\linewidth]{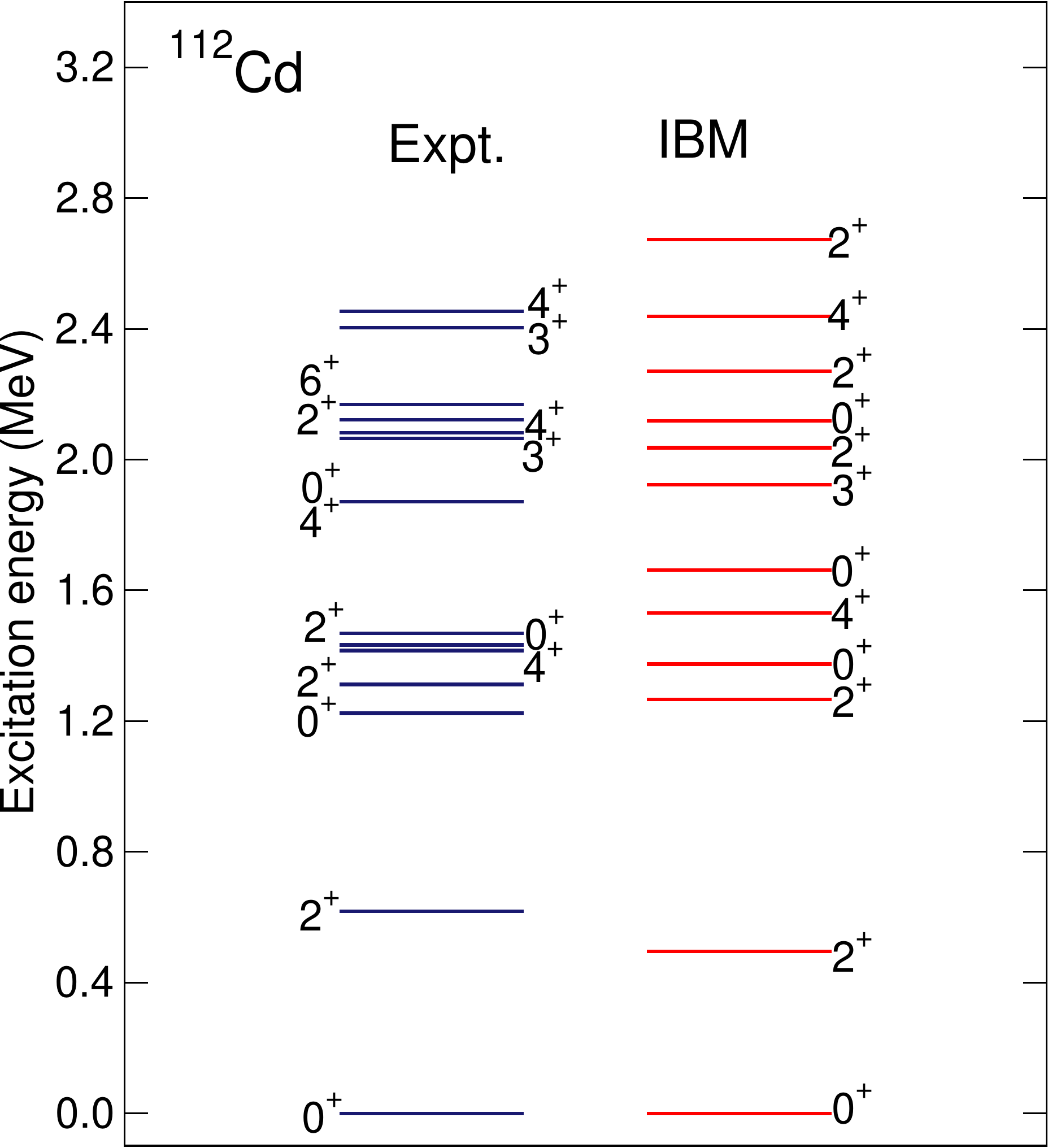} \\
\end{tabular}
\begin{tabular}{cc}
\includegraphics[width=0.28\linewidth]{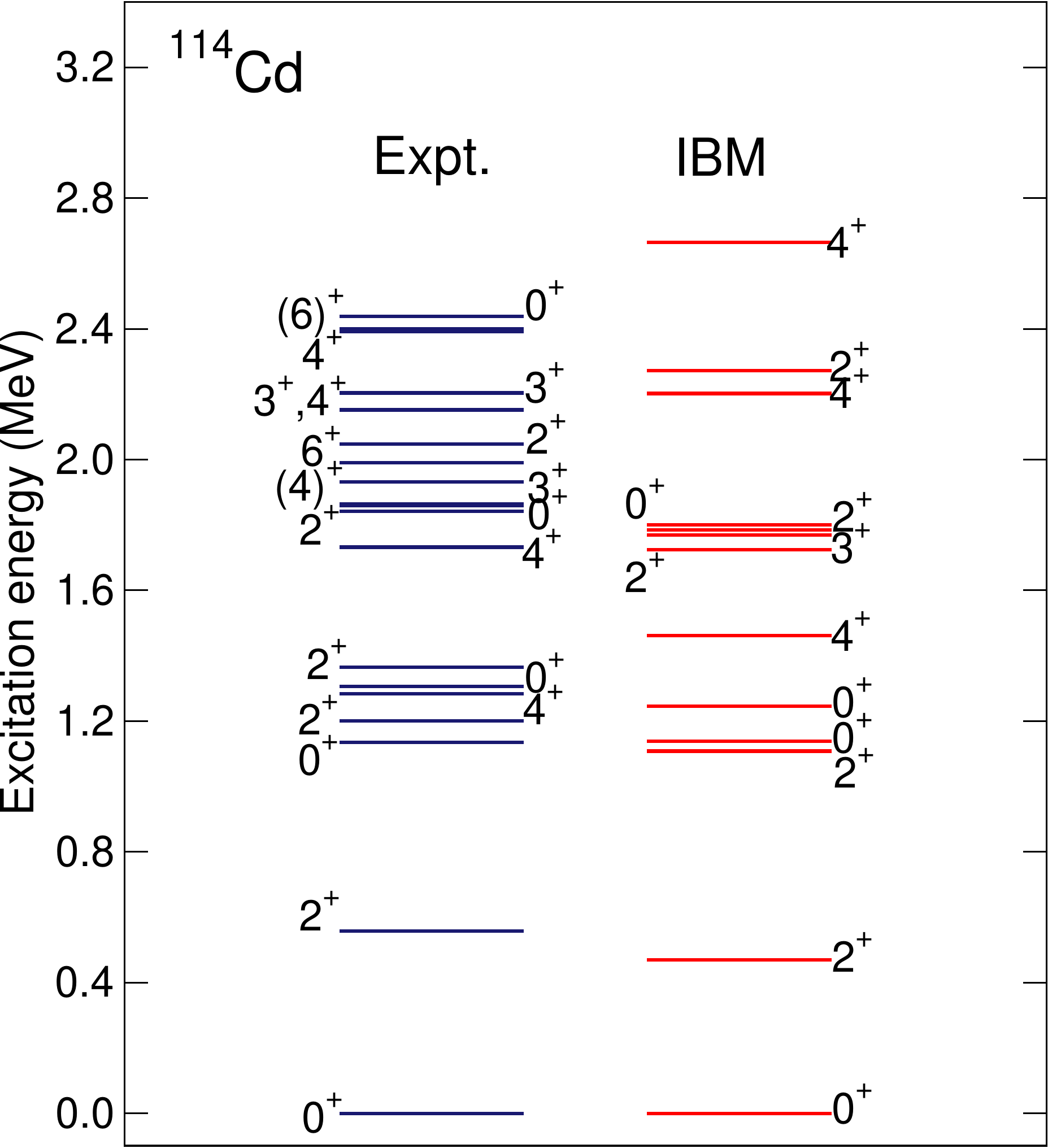} &
\includegraphics[width=0.28\linewidth]{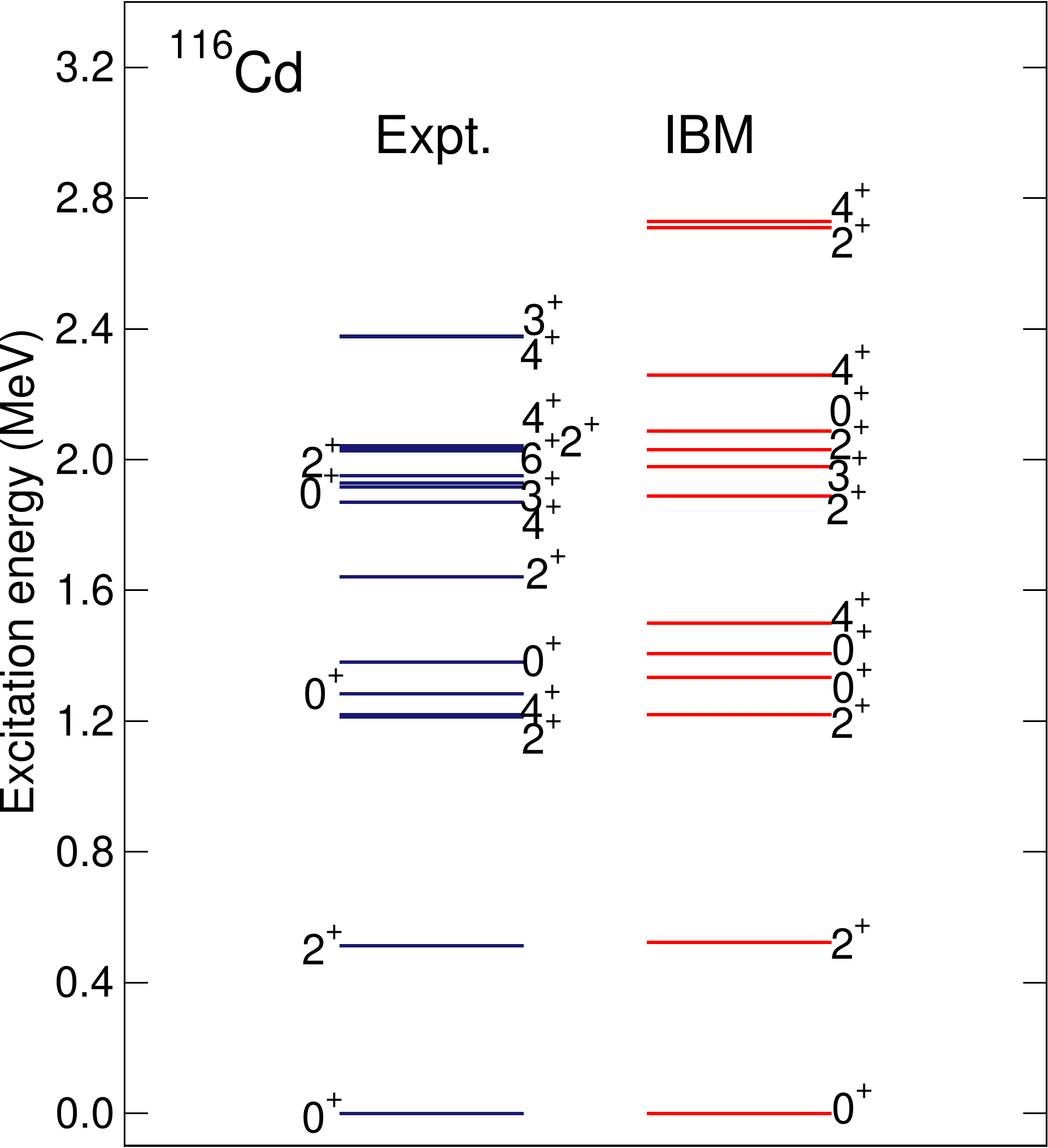} \\
\end{tabular}
\caption{(Color online) Experimental and predicted excitation spectra
 for the $^{108-116}$Cd nuclei. Based on Ref.~\cite{nomura2018cd}.} 
\label{fig:spectra}
\end{center}
\end{figure*}

Figure~\ref{fig:spectra} shows the comparison of the predicted and
experimental excitation spectra for $^{108-116}$Cd. 
The predicted energies of the intruder states can be found in those of the
third or higher 0$^{+}$ states. 
The SCMF-IBM calculation does predict the approximately right
excitation energies and the dependence on neutron number with the lowest
$0^+_3$ energy at the mid-shell nucleus $^{114}$Cd. 
The description of the spacing between both 0$^{+}$ states is
overestimated in $^{108,110}$Cd, but agrees quite well with experiment
in $^{112-116}$Cd. 
However, phenomenological calculations assign
for $^{114}$Cd the experimental $0^{+}_2$ state to the intruder
configuration. Therefore, the energy difference between
intruder and normal states is generally overestimated. 
The predicted normal states are systematically too 
deformed as can be seen by the energies of the first 2$^{+}$, 4$^{+}$
and especially 6$^{+}$ states. Experimentally states are observed with
more vibrational energies although there are serious problems with the
electric quadrupole transitions. 
The description of the energies of the 2$^{+}$ intruder state could be improved. 
Phenomenological calculations identified
in $^{110-114}$Cd the 2$^{+}_3$ state as the intruder state, while the
present calculation yield as main component the much higher lying
2$^{+}_4$ state. In contrast to the normal states the spacing between
the 0$^{+}$ and 2$^{+}$ intruder states is somewhat large.

We have also studied the $B(E2)$ transition rates. 
In view of the absence of fitting the data, there has been good agreement, and changing trends in the $B(E2)$ values have been 
well described \cite{nomura2018cd}. However, we found a few exceptions, mostly involving the
0$^{+}_{2}$, 0$^{+}_{3}$, 2$^{+}_{2}$ and 2$^{+}_{3}$ states as could be
expected from the different nature of these states compared to the
phenomenological IBM-2 calculations.  As an example, the $B(E2; 0^+_2
\rightarrow$ 2$^+_1$) are ten times underpredicted in $^{112,114}$Cd and
ten times overpredicted in $^{116}$Cd \cite{nomura2018cd}. This confirms that energies of these states,
which are the lowest where normal and intruder states mix, were not satisfactorily described. 
Another reason could be that at the SCMF level the prolate
minimum, from which the normal states are mainly constructed, was predicted to be
too deformed. 
Besides the $B(E2)$ transitions, an overall reasonable description of the $E0$ transition rates in the Cd nuclei has also been obtained (see, Ref.~\cite{nomura2018cd}).

\section{Octupole correlations in neutron-rich odd-mass Ba isotopes}
\label{sec-2}

Octupole deformation attracts a renowned interest in nuclear structure, and is also a declared objective of the RIB experiments. In addition, nuclei with octupole deformation are expected to show large electric dipole moment, which would imply the CP violation and thus points to physics beyond the standard model of elementary particles. In Ref.~\cite{nomura2018oct} we have extended the SCMF-IBM method to investigated the octupole correlations in neutron-rich odd-mass Ba isotopes in the mass $A\approx 140-150$ region,where octupole correlation is expected to be most pronounced. In this case, two additional degrees of freedom should be introduced to the $sd$-IBM system: (i) octupole ($f$) boson with spin and parity $J=3^-$; (ii) an unpaired nucleon (fermion) degree of freedom for describing odd-mass system. For the sake of simplicity, we do not distinguish neutron and proton degrees of freedom, and assume axial symmetry.

\subsection{Octupole deformation in even-even nuclei}

\begin{figure}[htb!]
\begin{center}
\includegraphics[width=0.8\linewidth]{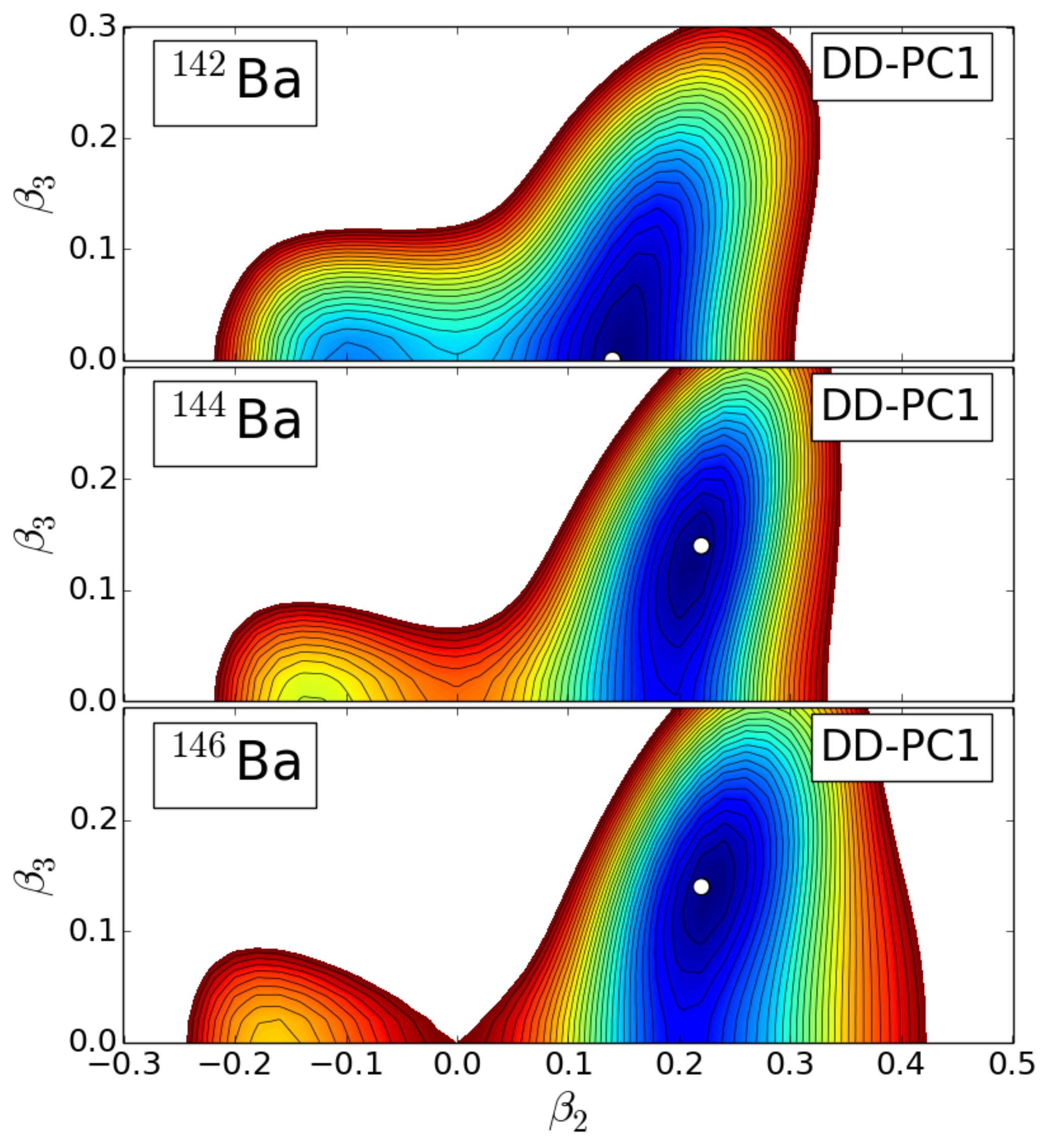}
\caption{(Color online) The SCMF ($\beta_{2}, \beta_{3}$) PESs for $^{142,144,146}$Ba, obtained with the DD-PC1 functional 
and a separable pairing force of finite range. The energy
 difference between neighbouring contours is 200 keV. Equilibrium minima
 are identified by open circles. Based on Ref.~\cite{nomura2018oct}.} 
\label{fig:pes-ba}
\end{center}
\end{figure}

\begin{figure}[htb!]
\begin{center}
\includegraphics[width=\linewidth]{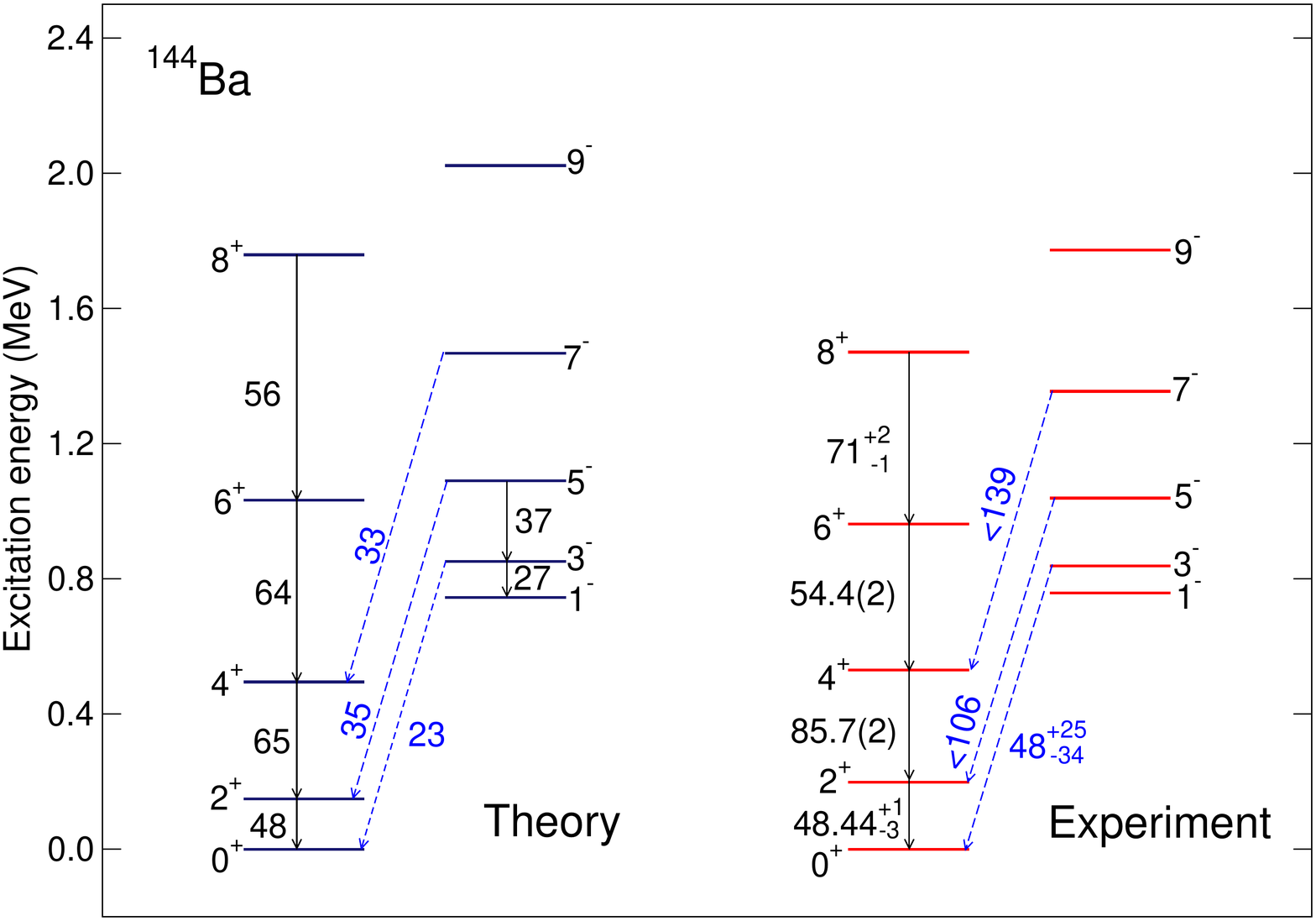}
\caption{(Color online) Low-energy positive- and negative-parity spectra
 of the even-even boson core nuclei $^{142,144,146}$Ba. The $B(E2)$
 (numbers along arrows within each band) and
 $B(E3)$ (inter-band, dashed arrows) values are given  
  in Weisskopf units. Experimental values are taken from Ref.~\cite{bucher2016}.} 
\label{fig:level-even}
\end{center}
\end{figure}

Figure~\ref{fig:pes-ba} depicts the PESs in terms of the axially-symmetric quadrupole $\beta_2$ and octupole $\beta_3$
deformations, computed by the constrained relativistic Hartree-Bogoliubov method \cite{vretenar2005}. 
For $^{142}$Ba the equilibrium minimum is found on the $\beta_3=0$ axis,
indicating that it has a weakly-deformed quadrupole shape. 
In $^{144,146}$Ba a minimum with non-zero $\beta_3$ deformation
($\beta\approx 0.1$) appears. The minimum is not very pronounced and is rather soft in
the $\beta_3$ direction, suggesting the occurrence of octupole vibrational states 
in these nuclei. 

The excitation spectra and transition rates for the octupole-deformed nuclei are computed by diagonalizing the IBM Hamiltonian comprising $J=0^+$ ($s$), 
$2^+$ ($d$), and  $3^-$ ($f$) boson degrees of freedom, which is determined from  
the SCMF ($\beta_{2},\beta_{3}$) PES. 
As an example, the low-energy level scheme for the $^{144}$Ba isotope is 
displayed in Fig.~\ref{fig:level-even}. In general, the theoretical predictions 
are in good agreement with the
experimental results \cite{bucher2016}, not only for the
excitation energies but also for the $E2$ and $E3$ transition
strengths. 
We note the very low-lying negative-parity band, which conforms to the results that the corresponding SCMF PES in Fig.~\ref{fig:pes-ba} exhibits a  rather pronounced $\beta_3\neq 0$ minimum, when compared to the one for $^{142}$Ba. 
A rather large value $B(E3;3^-\rightarrow 0^+)$ is
predicted for the $^{144}$Ba nucleus, but is still considerably smaller 
than the experimental value \cite{bucher2016}. 
Note, however, the large uncertainty of the latter.

\subsection{Particle-boson coupling}

In order to compute the spectroscopy of the odd-mass nuclei, we resort to the method developed in Ref.~\cite{nomura2016odd}, that is based on the SCMF calculation and the particle-core coupling. Here the even-even core is described by the IBM, and the interaction between the odd particle and the even-even boson core is modelled in terms of the interacting boson-fermion model (IBFM) \cite{IBFM}. 
The Hamiltonian of the IBFM is composed of the IBM Hamiltonian $\hat H_{\mathrm{B}}$, the Hamiltonian for a single fermion $\hat H_{\mathrm F}$, and the Hamiltonian $\hat H_{\mathrm{BF}}$ that couples fermion and boson spaces: 
\begin{eqnarray}
\hat H_\mathrm{IBFM}=\hat H_\mathrm{B}+\hat H_\mathrm{F}+\hat H_\mathrm{BF}. 
\end{eqnarray}
The single-fermion Hamiltonian $\hat H_\mathrm{F}=\sum_{j}\epsilon_j(-\sqrt{2j+1})(a_{j}^{\dagger}\tilde a_{j})^{(0)}$, where $\epsilon_j$ is the single-particle energy for the orbital $j$. 
For the sake of simplicity, we present the form of the Hamiltonian $\hat H_\mathrm{BF}$ within $sd$-IBFM. Based on the generalized seniority scheme, the following three terms of $\hat H_\mathrm{BF}$ have been shown to be most essential to the low-lying spectra in odd-mass nuclei \cite{IBFM}:  
\begin{equation}
\label{dyn}
\Gamma_0\sum_{jj^{\prime}}
\gamma_{jj^{\prime}}
\hat Q_B\cdot (a^{\dagger}_j\tilde a_{j^{\prime}})^{(2)}
\end{equation}
\begin{equation}
\label{exc}
-2\Lambda_0\sum_{jj^{\prime}j^{\prime\prime}}
\sqrt{\frac{5}{2j^{\prime\prime}+1}}\beta
_{jj^{\prime\prime}}\beta_{j^{\prime}j^{\prime\prime}}
:((d^{\dagger}\tilde a_{j})^{(j^{\prime\prime})}
(a^{\dagger}_{j^{\prime}}\tilde d)^{(j^{\prime\prime})})^{(0)}:
\end{equation}
\begin{equation}
\label{mon}
-A_0\sum_j\sqrt{2j+1}(a_j^{\dagger}\tilde a_{j})^{(0)}\hat n_d.
\end{equation}
The first, second and third terms are referred to as the dynamical
quadrupole, exchange, and monopole interactions, respectively. 
The quantities
$\gamma_{jj^{\prime}}=(u_ju_{j^{\prime}}-v_jv_{j^{\prime}})Q_{jj^{\prime}}$
and 
$\beta_{jj^{\prime}}=(u_jv_{j^{\prime}}+v_ju_{j^{\prime}})Q_{jj^{\prime}}$,
are given in terms of the occupation probabilities $u_j$ and $v_j$ 
for the orbital $j$ (satisfying  $u_j^2+v_j^2=1$) and
the matrix element of the quadrupole operator in the single-particle
basis $Q_{jj^{\prime}}=\langle j||Y^{(2)}||j^{\prime}\rangle$.
$\Gamma_0$, $\Lambda_0$, and $A_0$ denote strength parameters. 

The single-particle energies $\epsilon_j$ and occupation numbers $v^2_j$ are computed fully microscopically by the SCMF calculation constrained to zero deformation \cite{nomura2016odd}. The three strength parameters of $\hat H_\mathrm{BF}$, i.e., $\Gamma_0$, $\Lambda_0$, and $A_0$, are considered free parameters, and are fitted to reasonably reproduce experimental low-lying levels in a given nucleus. 
For the $sdf$-IBFM system, we have derived similar formulas to those in Eqs.~(\ref{dyn})--(\ref{mon}) from the generalized seniority scheme. More details can be found in Ref.~\cite{nomura2018oct}.

\subsection{Excitation spectra for the odd-mass nucleus}

\begin{figure}[h]
\begin{center}
\includegraphics[width=\linewidth]{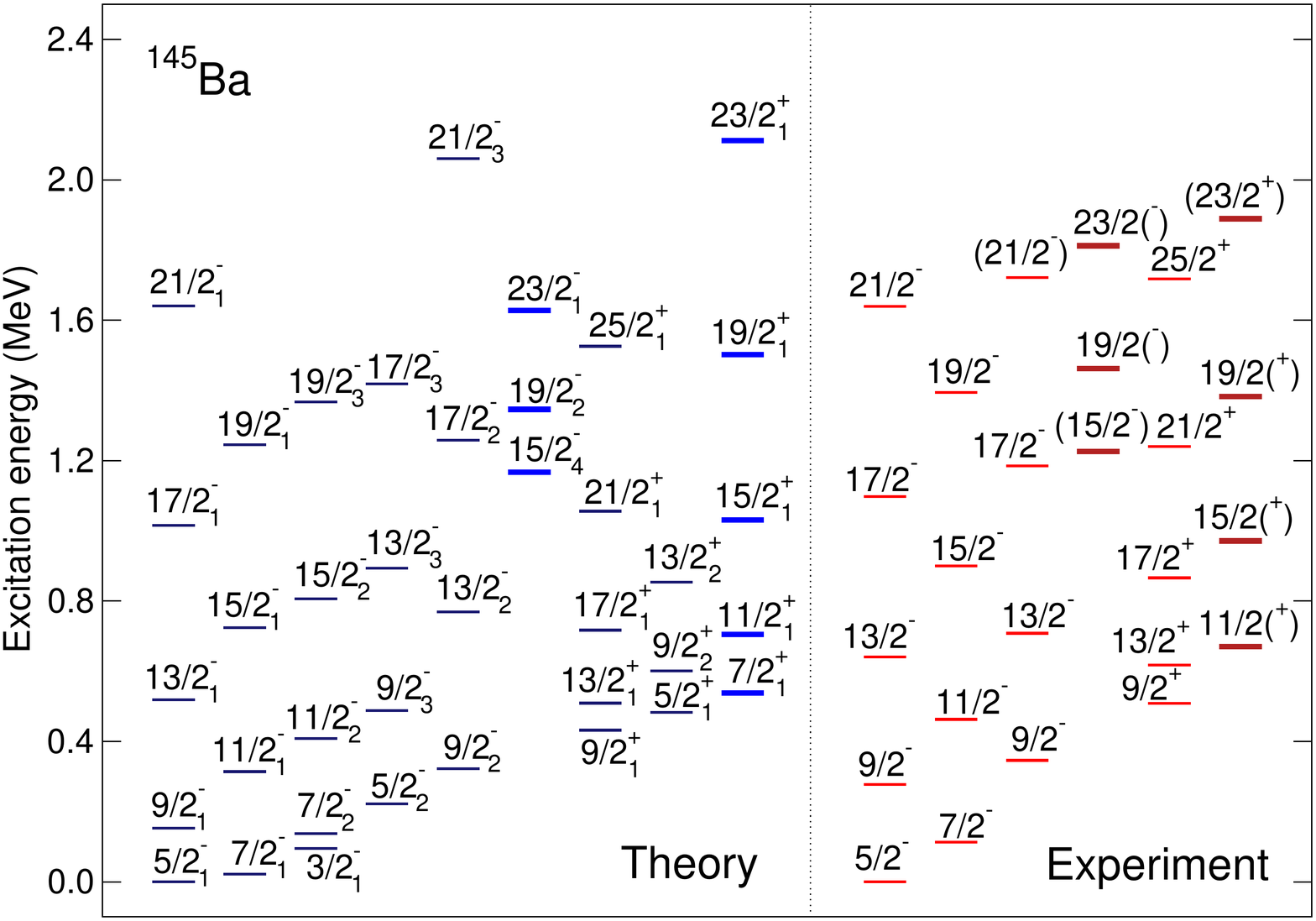}
\caption{(Color online) Theoretical and experimental \cite{rzacaurban2012} energy spectra for the nucleus
 $^{145}$Ba. Those theoretical levels that are suggested to contain one $f$ boson, as well as the experimentally suggested octupole bands, are marked as thick lines.} 
\label{fig:ba145-detail}
\end{center}
\end{figure}

Experimental information is available for the isotope $^{145}$Ba, that is enough to compare with the theoretical prediction. 
Since the corresponding even-even boson core nucleus 
$^{144}$Ba exhibits an octupole-soft potential at the SCMF level (cf. 
Fig.~\ref{fig:pes-ba}), we also expect that octupole correlations play an important role in the 
low-energy spectra of this odd-mass nucleus. The calculated excitation spectrum for $^{145}$Ba is compared to 
the corresponding experimental bands in Fig.~\ref{fig:ba145-detail}. 
The lowest two negative-parity bands in $^{145}$Ba are built on the 
${5/2}^-_1$ and ${7/2}^-_1$ states, which are characterized by the coupling of the 
unpaired neutron in the $1h_{9/2}$
single-particle orbital to the $sd$ boson space. 
The lowest positive-parity state ${9/2}^+_1$ is described
by the coupling of the $1i_{13/2}$ orbital to $sd$-boson states.

The theoretical negative-parity band
built on the ${15/2}^-_4$ state (calculated at 1167 keV) is dominated by 
the coupling of the $1i_{13/2}$ single-particle orbital to states with one
$f$-boson. Theoretically the ${15/2}^-_4$ state appears to be an
octupole state and is located close to the experimental
${15/2}^-$ state found at 1226 keV \cite{rzacaurban2012}. 
Moreover, rather strong E3 transitions from the $J^\pi={15/2}^-_4$ band to the
corresponding low-lying positive-parity bands are predicted: for
instance $B(E3; {15/2^-_4}\rightarrow {9/2^+_1})=25$ and  $B(E3; {19/2^-_2}\rightarrow
{13/2^+_1})=31$ (in W.u), comparable to 
the $B(E3; 3^-_1\rightarrow 0^+_1)$ value of 23 W.u. in the corresponding even-even
core nucleus $^{144}$Ba (see, Fig.~\ref{fig:level-even}). 
However, to verify model predictions, experimental information on 
the $B(E2)$ and $B(E3)$ values is needed.


\section{Summary}
I have presented a method of deriving the IBM Hamiltonian based on the SCMF framework. The constrained SCMF calculation with a given EDF is performed to obtain the PES in terms of the relevant collective coordinates. The SCMF PES is then mapped onto the expectation value of the IBM Hamiltonian in the boson coherent state. This procedure completely determines the strength parameters of the IBM, and the diagonalization of the mapped IBM Hamiltonian provides excitation spectra and transition probabilities. 
Two recent applications of the method are highlighted, that involve the configuration mixing of normal and intruder states, the octupole deformation, and the odd nucleon degree of freedom. 

As the EDF framework, on which the mapping procedure is built, provides a global and reliable mean-field description of intrinsic properties of nuclei, it has been made possible to determine the IBM Hamiltonian for arbitrary nuclei, starting only from the nucleonic degrees of freedom. This points to a crucial step toward the microscopic foundation of the IBM within a unified theoretical framework. 
In addition, both theoretical approaches, i.e., IBM and DFT, can be complementary to each other, in such a way that  the latter gives microscopic input with the former, and the former facilitates spectroscopic calculations, which are, from both computational and methodological points of view, highly demanding with the latter approach. 
On the practical side, contrary to the conventional IBM studies, using the present method we have gained the capability of predicting the spectroscopic properties of nuclei in an accurate, systematic, and mathematically simple way. This opens up a new possibility to study heavy exotic nuclei that are now under rigorous investigations at RIB facilities around the world. 

\begin{acknowledgement}
The results reported here are based on the works with J. Jolie, T. Nik\v si\'c, T. Otsuka, N. Shimizu, and D. Vretenar. 
This work is financed within the Tenure Track Pilot Programme of the Croatian Science Foundation and the \'Ecole Polytechnique F\'ed\'erale de Lausanne and the Project TTP-2018-07-3554 Exotic Nuclear Structure and Dynamics, with funds of the Croatian-Swiss Research Programme. 
\end{acknowledgement}

%

\begin{thebibliography}{99}
%
%
\bibitem{IBM}
F. Iachello, and A. Arima, \textit{The interacting boson model} (Cambridge University Press, Cambridge, 1987)

\bibitem{OAI}
T. Otsuka, A. Arima, and F. Iachello, Nucl. Phys. A \textbf{309}, 1 (1978)

\bibitem{nomura2008}
K. Nomura, N. Shimizu, and T. Otsuka, Phys. Rev. Lett. \textbf{101}, 142501 (2008)

\bibitem{nomura2011rot} 
K. Nomura et al., Phys. Rev. C \textbf{83}, 041302 (2011)

\bibitem{nomura2012tri}
K. Nomura, N. Shimizu, D. Vretenar, T. Nik\v si\'c, and T. Otsuka, Phys. Rev. Lett. \textbf{108}, 132501 (2012)

\bibitem{nomura2014} K. Nomura et al., Phys. Rev. C \textbf{89}, 024312 (2014)


\bibitem{nomura2016sc} 
K. Nomura et al.,  J. Phys. G \textbf{43}, 024008 (2016)


\bibitem{nomura2018cd} K. Nomura, and J. Jolie, Phys. Rev. C \textbf{98}, 024303 (2018)

\bibitem{nomura2018oct} K. Nomura, T. Nik\v si\'c, and D. Vretenar, Phys. Rev. C \textbf{97}, 024317 (2018). 


\bibitem{vretenar2005}
D. Vretenar et al., Phys. Rep. \textbf{409}, 101 (2005)

\bibitem{bender2003}
M. Bender et al., Rev. Mod. Phys. \textbf{75}, 121 (2003) 

\bibitem{BM}
A. Bohr, and B. R. Mottelson, \textit{Nuclear Structure} (Benjamin, N.Y., 1975)

\bibitem{ginocchio1980} J. N. Ginocchio and M. W. Kirson, Nucl. Phys. A \textbf{350}, 31 (1980).


\bibitem{duval1981}
P. D. Duval and B. R. Barrett, Phys. Lett. B \textbf{100}, 223 (1981).

\bibitem{frank2004}
A. Frank et al., Phys. Rev. C \textbf{69}, 034323 (2004).





\bibitem{bucher2016} B. Bucher et al., Phys. Rev. Lett. \textbf{116}, 112503 (2016)

\bibitem{nomura2016odd}
K. Nomura et al., Phys. Rev. C \textbf{93}, 054305 (2016). 

\bibitem{IBFM}
F. Iachello, and P. Van Isacker, \textit{The Interacting Boson-Fermion
Model} (Cambridge University Press, Cambridge, England,
1991).

\bibitem{rzacaurban2012}
T. Rzaca-Urban et al., Phys. Rev. C \textbf{86}, 044324 (2012).



\end{thebibliography}
%
%

\end{document}